\documentstyle[twocolumn,aps,psfig]{revtex}
\newcommand{\ignore}[1]{}
\begin{document}
\def\be{\begin{equation}}
\def\ee{\end{equation}}
\def\ba{\begin{eqnarray}}
\def\ea{\end{eqnarray}}

\draft

\title{Wave Function Structure in Two-Body Random Matrix Ensembles}
\author{Lev Kaplan\thanks{lkaplan@phys.washington.edu} 
 and Thomas Papenbrock\thanks{papenbro@phys.washington.edu}
 \\Institute for Nuclear Theory and Department of Physics,\\
University of Washington, Seattle, WA 98195
}
\maketitle

\begin{abstract}
We study the structure of eigenstates in two-body interaction random matrix
ensembles and find significant deviations from random matrix theory
expectations. The deviations are most prominent in the tails of the
spectral density and indicate localization of the eigenstates in Fock
space. Using ideas related to scar theory we derive an analytical formula
that relates fluctuations in wave function intensities to fluctuations of
the two-body interaction matrix elements. Numerical results for many-body
fermion systems agree well with the theoretical predictions.
\end{abstract}

Random matrix theory (RMT) has become a powerful tool for
describing statistical properties of wave functions and energy levels in
complex quantum systems~\cite{Brody81,GMW98}. The use of a two-body random
matrix ensemble (TBRE)~\cite{FW70,BF71} is of particular interest for many-body
systems since classical RMT implies the presence of $k$-body forces ($k>2$) and
gives the unphysical semicircle as the spectral density. The TBRE displays the
same spectral fluctuations as classical RMT while its Gaussian spectral density
agrees well with nuclear shell model calculations. The situation is
not so clear for the structure
of wave functions in TBRE. Recent results show that ground states of
shell model Hamiltonians with two-body random interactions favor certain
quantum numbers~\cite{BJ98} and thereby differ considerably from RMT
expectations. 

In this letter we examine the structure of wave functions in TBRE, and compare
numerical results with theoretical predictions. Such a
study is not only interesting on its own but is also motivated by the ongoing
importance of TBRE for nuclear~\cite{BJ98,Kota} and mesoscopic physics
\cite{Shepel,Silvestrow}. We recall that deviations from RMT indicate some
degree of wave function non-ergodicity and are related to phenomena like 
Fock space localization in many-body systems \cite{Altshuler,HAW}, and scars of
periodic orbits \cite{Heller,Kaplan} or invariant manifolds \cite{manfold} in
classically chaotic systems.

To quantify the degree of localization of a given wave function, it is useful
to introduce the notion of an inverse participation ratio (IPR). Thus, let $D$
be the total dimension of the relevant Fock subspace, let $|b\rangle$ label the
single-particle basis states (Fock states), and let $|\alpha\rangle$ represent
the eigenstates of the Hamiltonian. Then the overlap intensities
\begin{equation}
P_{\alpha b}= |\langle \alpha|b \rangle|^2
\end{equation}
are the squares of the expansion coefficients, and have mean value $1/D$.
The IPR of eigenstate $|\alpha\rangle$
is defined as the first nontrivial moment of the intensity
distribution, namely the ratio of the mean squared
$P_{\alpha b}$ to the square of the mean:
\begin{equation}
\label{ipr}
{\rm IPR}_\alpha = {{1 \over D} \sum_{b=1}^D P_{\alpha b}^2 \over
\left ( {1 \over D} \sum_{b=1}^D P_{\alpha b} \right )^2}
= D\sum_{b=1}^D P_{\alpha b}^2  \,.
\end{equation}
The IPR measures the inverse fraction of Fock states that participate in
building up the full wave function $|\alpha\rangle$, i.e. ${\rm IPR}_\alpha =1$
for a wave function that has equal overlaps $P_{\alpha b}$ with all basis
states, and ${\rm IPR}_\alpha=D$ for the other extreme of a wave function that
is composed entirely of one basis state. While complete information about
wave function ergodicity is contained in the full
distribution of intensities
$P_{\alpha b}$, the IPR serves as a very useful one-number measure of the
degree of Fock-space localization.  In RMT, the $P_{\alpha b}$ are given (in
the large-$D$ limit) by squares of Gaussian random variables, in accordance
with the Porter-Thomas distribution, leading to ${\rm
IPR}_{\rm RMT}=3$ for real wave functions. For finite $D$, the IPR is slightly
below its asymptotic value (with the deviation falling off as $1/D$), but is
still uniform over the entire spectrum.

A very simple two-body interaction model, however, already displays behavior
which is qualitatively different from this naive RMT expectation.  Let the
Hamiltonian be given by
\begin{equation}
\label{h1}
H={1\over 2}\sum_{i,j,k,l} V_{ijkl} a^\dagger_i a^\dagger_j a_k a_l \,,
\end{equation}
where the single-particle indices $i$, $j$, $k$, $l$ run from $1$ to $M$ (the
number of available single-particle states) and the $V_{ijkl}$ are Gaussian
random variables with unit variance. The operators $a^\dagger_j$ and $a_j$
create and annihilate a fermion in the single-particle state labeled by $j$,
respectively, and obey the usual anti-commutation rules. The dimension of the
$N$-particle Fock subspace is given by $D={M \choose N}$.  We notice that this
model contains no explicit one-body terms and therefore
does not display Anderson-type
localization effects. On physical grounds, we assume that the two-body
interaction is generated by a real symmetric potential, $V(r_1,r_2)=
V(r_2,r_1)$ and choose real single-particle wave functions.  These conditions
lead to the constraints $V_{ikjl}=V_{ljki}=V_{jilk}=V_{ijkl}$ and reduce the
number of independent variables describing
a given realization (Eq.~\ref{h1}) of the
ensemble. While such correlations between matrix elements will determine
factors of $2$ in the calculation below,
they do not qualitatively affect our results. The choice of
bosons rather than fermions also does not qualitatively change the localization
behavior.

\begin{figure}
\centerline{
\psfig{file=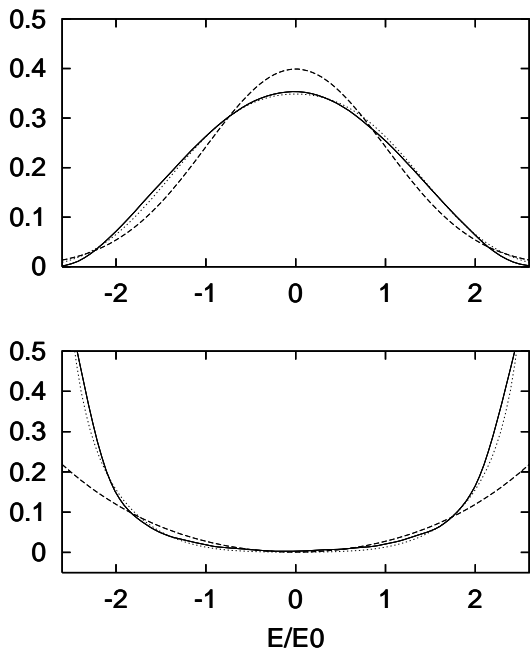,angle=270,width=3in}}
\vskip 0.1in
\caption{Spectrum and IPR for $N=6$
fermions distributed over $M=13$ orbitals. Data is averaged over
an ensemble of $20$ systems of the form (Eq.~\ref{h1}).
Top: Normalized spectrum
$\rho(E)/D$,
Eq.~\ref{rhoe} (solid curve), Gaussian prediction, Eq.~\ref{gauss}
(dashed curve), and modified Gaussian, Eq.~\ref{modgauss} (dotted curve).
Bottom: ${\rm IPR}/{\rm IPR}_{\rm RMT}-1$: data (solid curve), simple theory,
Eq.~\ref{iprh1} (dashed curve) and theory based on modified Gaussian
form (dotted curve).
}
\label{fig1}
\end{figure}

Fig.~\ref{fig1}(top) shows a smoothed ensemble-averaged spectral density
\begin{equation}
\label{rhoe}
\rho(E)={\rm Tr} \; \delta(E-H)=\sum_{\alpha=1}^D \delta(E-E_\alpha) \,,
\end{equation}
for $M=13$ and $N=6$, while in Fig.~\ref{fig1}(bottom) we plot
${\rm IPR}/{\rm IPR}_{\rm RMT}-1$ as a
function of energy. Obviously, we observe strong deviations from RMT behavior
at the edges of the spectrum. This implies that wave function intensities
do not have a Porter-Thomas distribution. 

To understand this surprising result, we may adapt a formalism previously
used successfully to understand wave function scars and other types of
anomalous quantum localization behavior in single-particle systems.
Let $\rho_b(E)$ be the local density of states (strength function)
of the basis state $b$:
\begin{equation}
\label{rhob}
\rho_b(E)=\langle b|\delta(E-H)|b\rangle =\sum_{\alpha=1}^D P_{\alpha b}
\delta(E-E_\alpha) \,;
\end{equation}
it is given by the Fourier transform of the autocorrelation function
\begin{equation}
A_b(t)=\langle b |e^{-iHt}|b\rangle \,.
\end{equation}
Let us assume that $A(t)$ displays only two different
time scales: the initial decay time $T_{\rm decay}$ of the Fock
state $|b\rangle$ due to interactions, and the Heisenberg time
$T_H \sim D T_{\rm decay}$ (i.e. $\hbar$ over the mean level spacing)
at which individual eigenlevels are resolved. Following the initial
decay, random long-time recurrences in $A_b(t)$ can be shown to be convolved
with the short-time behavior~\cite{shorttime,Kaplan}. In the energy domain
this produces random oscillations $f_b(E)$ multiplying a smooth envelope
$\rho_b^{\rm sm}(E)$ given
by the short-time dynamics:
\begin{equation}
\rho_b(E) = f_b(E) \rho_b^{\rm sm}(E)\,.
\end{equation}
Here 
\begin{equation}
\label{sum}
\sum_{b=1}^D \rho_b^{\rm sm}(E) =\rho^{\rm sm}(E) \,,
\end{equation}
while $f_b(E)$ is a fluctuating function with mean value of unity:
\begin{equation}
f_b(E)= {\sum_{\alpha=1}^D r_{\alpha b} \delta(E-E_\alpha) \over
\rho^{\rm sm}(E)}\,.
\end{equation}
The $r_{\alpha b}$ are random $\chi^2$ variables with mean value one.
Then, substituting into Eq.~\ref{rhob},
we have the individual wave function intensities given by
\begin{equation}
P_{\alpha b}=r_{\alpha b} {\rho_b^{\rm sm}(E_\alpha) \over
\rho^{\rm sm}(E_\alpha)}\,.
\label{pabprod}
\end{equation}
Both $r_{\alpha b}$ and $\rho_b^{\rm sm}$ have $b$-dependent fluctuations,
which are uncorrelated under our assumption of separated time scales. Then
using Eq.~\ref{pabprod} we can express the IPR
(Eq.~\ref{ipr})
as ($\delta\rho_b^{\rm sm}(E)\equiv\rho_b^{\rm sm}(E)-
\langle\rho_b^{\rm sm}(E)\rangle$) 
\begin{eqnarray}
\label{iprproduct}
{\rm IPR}_\alpha &=& {\langle r_{\alpha b}^2 \rangle \over
\langle  r_{\alpha b} \rangle^2}  \times
{\langle \rho_b^{\rm sm}(E_\alpha)^2 \rangle \over
\langle \rho_b^{\rm sm}(E_\alpha) \rangle^2} \nonumber \\
& = & {\rm IPR}_{\rm RMT} 
\left ( 1+ {\langle \delta \rho_b^{\rm sm}(E_\alpha)^2 \rangle \over
\langle \rho_b^{\rm sm}(E_\alpha) \rangle^2} \right ) \,,
\end{eqnarray}
where all averages are over the Fock basis index $b$, i.e. 
$\langle \ldots \rangle \equiv 
D^{-1} \sum_{b=1}^D$.

In the limit of many particles (and many holes) $N$, $M-N \gg 1$ 
the spectrum approaches a Gaussian shape~\cite{Gervots,Mon}
\begin{equation}
\label {gauss}
\rho^{\rm sm}(E) = {D \over \sqrt{2 \pi E_0^2}}
\exp{(-E^2/2 E_0^2)}\,,
\end{equation}
where $E_0^2=D^{-1} {\rm Tr} \; H^2$ is given by the mean sum of
squares of matrix
elements in any given row of the Hamiltonian. The same arguments lead to
the vanishing of higher-order cumulants for the individual strength
functions $\rho_b^{\rm sm}(E)$, so each of
these should also have a Gaussian
shape, but with centroid $c_b=H_{bb}$ and variance $v_b=\sum_{b'} H_{bb'}^2$. 
Due to these fluctuations in the centroids and widths as one goes through
the different basis states $b$, we have
\begin{eqnarray}
\label{drho}
\delta \rho_b^{\rm sm}(E) &=& {\partial \rho_b^{\rm sm}(E) \over \partial c_b}
\delta c_b + {\partial \rho_b^{\rm sm}(E) \over \partial v_b} \delta v_b \nonumber \\
&=& \left [{E \over E_0} {\delta c_b \over E_0}
+ \left(
{E^2 \over E_0^2} -1 \right ) {\delta v_b \over 2 E_0^2}
\right ] \rho^{\rm sm}(E)
\,,
\end{eqnarray}
where in the second line we have used the Gaussian form of Eq.~\ref{gauss}.
Substituting into Eq.~\ref{iprproduct}, we obtain the general form:
\begin{equation}
\label{iprgeneral}
{{\rm IPR}(E) \over {\rm IPR}_{\rm RMT}}-1 = 
{\langle (\delta c_b)^2 \rangle
\over E_0^2} \left ( {E^2 \over E_0^2} \right ) +
{\langle (\delta v_b)^2 \rangle
\over 4 E_0^4} \left ( {E^2 \over E_0^2} -1 \right )^2 \,,
\end{equation}
valid of course not only for two-body interactions but for any ergodic
Hamiltonian with a Gaussian density of states. The quantities $\langle (\delta
c_b)^2 \rangle$ and $\langle (\delta v_b)^2\rangle$ do depend on the
parameters of the model, but remarkably the IPR behavior is always given to
leading order by a quartic polynomial in energy. From this point of view, it is
easy to understand the enhancement in fluctuations near the edge of the
spectrum (i.e. for large $|E|$): two Gaussians differing only slightly in their
centroid or width may look almost identical in the bulk, but the relative
difference increases dramatically as one moves into the tail of the spectrum.

For our simple model (Eq.~\ref{h1}) one may easily compute the coefficients in
Eq.~\ref{iprgeneral}. The squared width $\langle v_b \rangle = E_0^2$ of the
full spectrum is given by the sum of squares of entries in one row of the
Hamiltonian and equals the number of independent terms in Eq.~\ref{h1}
that couple any
given Fock state to other Fock states, times the mean squared value ($=1$)
of each
such term. A simple counting argument then shows
$E_0^2=2{N \choose 2} \left [1+2(M-N)+{M-N \choose 2}\right ]$. Because all
the contributions are independent $\chi^2$ variables of mean $1$ and
variance $2$, the
variance in the sum is given by twice the number of contributions:
$\langle(\delta v_b)^2 \rangle= 2E_0^2$. Finally, the variance in the centroid
is given by the number of terms in the Hamiltonian
(Eq.~\ref{h1}) which contribute
to each diagonal element in the Fock basis, namely $\langle(\delta c_b)^2
\rangle={N \choose 2}$. Then
\begin{eqnarray}
{{\rm IPR}(E) \over {\rm IPR}_{\rm RMT}}-1 &=&
\left [ 1+2(M-N)+{M-N \choose 2} \right ]^{-1} \nonumber \\ 
\label{iprh1} & \times &
\left [\left ( {E^2 \over E_0^2} \right ) +
\left [ 2 {N \choose 2} \right ]^{-1}
\left ( {E^2 \over E_0^2} -1 \right )^2 \right ] \\
& \approx & {2 \over M^2} \left [{E^2 \over E_0^2}+
{1 \over N^2} \left ( {E^2 \over E_0^2} -1 \right )^2 \right ] \,,
\end{eqnarray}
where in the last line we have taken the dilute many-particle limit
$M \gg N \gg 1$. The fluctuations are always strongest near the edge
of the spectrum; in particular close to the ground state 
($E_{\rm gs}^2 \approx 2 E_0^2 \ln D \approx 2 E_0^2 N \ln (M/N)$),
we find
\begin{equation}
{{\rm IPR}(E_{\rm gs}) \over {\rm IPR}_{\rm RMT}}-1=
{4 \over M^2} \left [ N \ln {M \over N}+ 2 \ln^2 {M \over N} \right ] \,.
\end{equation}

Let us compare the IPR prediction of Eq.~\ref{iprh1} with the numerical data
presented in Fig.~\ref{fig1}(bottom).  While the prediction
qualitatively describes the correct trend of the IPR as a function of energy,
it is not in
quantitative agreement with the data and fails by as much as a factor
of $2$ at the edge of the spectrum.  This is not very surprising, given that
the actual behavior of the spectral density $\rho^{\rm sm}(E)$ is far from
Gaussian for the parameters we have chosen. Although it is known that the
spectrum approaches a Gaussian form in the many-particle
limit~\cite{Gervots,Mon}, the number of particles that can be simulated
numerically is not nearly large enough for the Gaussian form to be a good
quantitative approximation, particularly in the tail
(see Fig.~\ref{fig1}(top)). A
good way to measure deviations from the Gaussian form is to compute the fourth
cumulant divided by the square of the second cumulant $E_0^2$: for a
Gaussian this quantity vanishes while for our system it equals $-0.7$,
actually closer to the semicircle value of $-1$.

To obtain quantitatively valid predictions, we must correct for deviations from
the Gaussian shape. The ansatz 
\be 
\rho^{\rm sm}(E)={1-6\epsilon \left({E\over
E_0}\right)^2+\epsilon \left({E\over E_0}\right)^4\over 1-3\epsilon} {D \over
\sqrt{2 \pi E_0^2}} \exp{\left({-E^2\over 2 E_0^2}\right)} 
\label{modgauss}
\ee 
allows for nonzero higher cumulants while keeping the number of states and the
width $E_0^2$ fixed. The parameter $\epsilon$ is obtained by a least squares
fit of the numerical data to this form. A derivation starting from the
first line of 
Eq.~\ref{drho} yields an IPR that depends on the energy through a rational
function instead of the quartic polynomial of Eq.~\ref{iprh1}. As we can see
in Fig.~\ref{fig1}(bottom),
this leads to a surprisingly good quantitative prediction
for the IPR behavior, given that we are applying perturbation theory around a
Gaussian shape for a spectrum that in reality is very far from the Gaussian
limit.

Gaussian spectral densities have been observed in nuclear shell model
calculations with realistic \cite{Zelevinsky} or random two-body
interactions \cite{FW70,BF71}. This is mainly due to the presence of spin
and isospin degrees of freedom, 
which cause important correlations between Hamiltonian matrix
elements. Let us therefore consider adding spin to the Hamiltonian
(Eq.~\ref{h1}), so that $a^\dagger_j$ and $a_j$ now create
and annihilate a
fermion in the single-particle state labeled by $j\equiv (n_j,s_j)$,
with $n_j$ and $s_j=\pm 1/2$ denoting the orbital and spin quantum numbers.
In what follows we consider $N$ fermions in a shell of fixed total spin
$S=\sum_{j=1}^N
s_j=0$. We assume that the random matrix elements $V_{ijkl}$
depend only on the orbital quantum numbers; this reduces the number of
independent matrix elements considerably. The (sparse) two-body
interaction matrix is constructed
using a code similar to the one described in Ref.\cite{BEC}.

\begin{figure}
\centerline{
\psfig{file=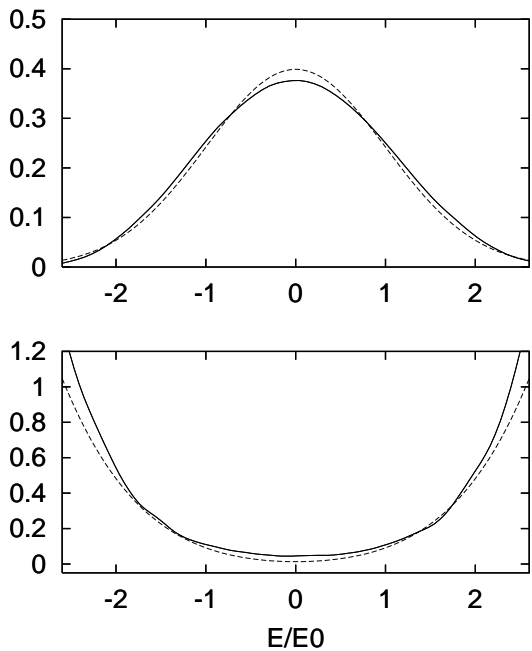,angle=270,width=3in}}
\vskip 0.1in
\caption{Spectrum and IPR for $N=6$
fermions with spin distributed over $M=7$ orbitals. Data is averaged over
an ensemble of $20$ systems of the form (Eq.~\ref{h1}).
Top: Normalized spectrum
$\rho(E)/D$,
Eq.~\ref{rhoe} (solid curve) and Gaussian prediction, Eq.~\ref{gauss}
(dashed curve).
Bottom: ${\rm IPR}/{\rm IPR}_{\rm RMT}-1$: data (solid curve) and theory,
Eq.~\ref{iprgeneral} (dashed curve).
}
\label{fig2}
\end{figure}

The spectral density of this TBRE agrees very well with a Gaussian, and the IPR
is predicted by Eq.~\ref{iprgeneral}.
However, the determination of the quantities
$\langle (\delta c_b)^2 \rangle$ and $\langle (\delta v_b)^2\rangle$ is more
difficult since the spin degree of freedom makes the required
counting of matrix elements a non-trivial task. Alternatively, one may obtain
these quantities directly from the numerically
generated Hamiltonians. Fig.~\ref{fig2} shows that the numerical data and
theoretical results are in good agreement. This confirms the validity of the
simple theory derived in this letter. 

We note
that our results are in qualitative agreement with nuclear shell model
calculations using realistic interactions
\cite{Zelevinsky,Whitehead78,Verb,BB} and with calculations in atomic physics
\cite{Flambaum}. Our calculation shows that deviations from ergodicity
near the ground state do not require the presence of one-body terms
in the Hamiltonian, though such terms may of course enhance the degree
of localization.

In summary we have studied numerically and analytically the wave function
structure in many-body fermion systems with random two-body
interactions. Near the edge of the spectrum,
wave function intensities of this
two-body random ensemble exhibit fluctuations that deviate strongly from random
matrix theory predictions, while good
agreement is obtained in the bulk of the spectrum. The numerical results agree
well with the theoretical prediction that is derived from arguments used in
scar theory. In particular, we have presented a simple formula that relates
fluctuations of the wave function intensities to fluctuations of the
two-body matrix elements of the Hamiltonian.

We thank George F. Bertsch for suggesting this study and for several very
useful discussions. This research was supported by the DOE under Grant
DE-FG-06-90ER40561.

\end{document}